\documentclass[twocolumn,nofootinbib,longbibliography,prl,superscriptaddress]{revtex4-1}

\usepackage{feynmp}
\usepackage{amsmath}
\usepackage{graphicx}
\usepackage{multirow}
\usepackage{latexsym}
\usepackage{textcomp}
\usepackage{verbatim}
\usepackage{color}
\usepackage{bm}
\usepackage{subfigure}
\usepackage{everysel}
\usepackage{keyval}
\usepackage{ragged2e}
\usepackage{dsfont}
\usepackage{amssymb}
\usepackage{enumitem}
\usepackage{pstricks}
\usepackage{mathtools}
\usepackage{braket}
\usepackage{slashed} 
\usepackage[colorlinks=true]{hyperref} 
\usepackage{tikz}

\usepackage{MnSymbol}

\usepackage{graphicx}
\usepackage{xcolor}
\usepackage{amsmath}
\usepackage{amsfonts}
\usepackage{bbm}

\usepackage{ulem}

\newcommand{\osum}{{%
    \setbox0\hbox{\circ}%
    \rlap{\hbox to \wd0{\hss\sum\hss}}\box0
}}

\newcommand{\tr}{{\textrm{tr}}}
\newcommand{\Tr}{{\textrm{Tr}}}

\newcommand{\cG}{{\mathcal{G}}}
\newcommand{\cH}{{\mathcal{H}}}
\newcommand{\cL}{{\mathcal{L}}}
\newcommand{\cO}{{\mathcal{O}}}

\newcommand{\cD}{{\mathcal{D}}}
\newcommand{\cJ}{{\mathcal{J}}}

\newcommand{\cN}{{\mathcal{N}}}
\usepackage{xr}

\begin{document}

\title{Unitarity of Symplectic Fermion in $\alpha$-vacua with Negative Central Charge }

\author{Shinsei Ryu}
\thanks{Electronic Address: shinseir@princeton.edu}
\affiliation{Department of Physics, Princeton University, Princeton, New Jersey, 08544, USA}

\author{Junggi Yoon}
\thanks{Electronic Address: junggi.yoon@apctp.org}
\affiliation{Asia Pacific Center for Theoretical Physics, POSTECH, Pohang 37673, Korea}
\affiliation{Department of Physics, POSTECH, Pohang 37673, Korea}

\date{\today}

\begin{abstract} 
We study the two-dimensional free symplectic fermion theory with anti-periodic boundary condition. 
This model has negative norm states with a naive inner product. 
This negative norm problem can be cured by introducing a new inner product. 
We demonstrate that this new inner product follows from the connection between the path integral formalism 
and the operator formalism. This model has a negative central charge, $c=-2$, and we clarify how CFT$_2$ 
with negative central charge can have a non-negative norm. 
Furthermore, we introduce $\alpha$-vacua in which the Hamiltonian is seemingly non-Hermitian. 
In spite of non-Hermiticity, we find that the energy spectrum is real. 
We also compare a correlation function with respect to the $\alpha$-vacua with that of the de Sitter space.
\end{abstract}

\maketitle

\textit{Introduction.}---Unitarity, as a primary postulate of quantum theory, 
has played a central role in producing momentous achievements in physics 
such as the optical theorem for the $S$-matrix and the Page curve in the black hole~\cite{Page:1993wv}. 
On the other hand, an open quantum system, which features non-unitarity, 
has been recently spotlighted. 
There, the physical properties of non-unitary models have been extensively investigated, 
such as topological phases~\cite{Rudner:2009ta,Esaki:2011uj,Sato:2011yw,Hu:2011qk,Kawabata:2020olo} 
and phase transitions~\cite{Kawabata:2017gkk,Xiao:2019vc,Dora:2018flx,Xiao:2021wo,Kawabata:2022biv} of non-Hermitian systems.

In two-dimensional conformal field theory (CFT$_2$), 
the central charge is known to be a powerful criterion for unitarity; 
a CFT$_2$ with negative central charge has negative norm states and therefore is non-unitary. 
CFTs with negative central charge such as 
the Yang-Lee edge singularity~\cite{Yang:1952be,Lee:1952ig,Fisher:1978pf,Cardy:1985yy,Itzykson_1989} 
and the PT-symmetric Su-Schrieffer-Heeger model~\cite{Chang:2019jcj} have provided fruitful laboratories 
to understand non-unitary physics.

The symplectic fermion theory--i.e., the anti-commuting scalar field theory---
has been thoroughly studied as an example of CFT$_2$ with central charge $c=-2$~\cite{Kausch:1995py,Kausch:2000fu} 
and as an example of a logarithmic CFT for the case of periodic boundary condition~\cite{Gaberdiel:1998ps}. 
It was proposed in~\cite{LeClair:2007iy,Robinson:2009xm} that a new inner product can cure 
the negative norm states of the symplectic fermion; 
therefore this model with the new inner product was claimed to be unitary~\cite{LeClair:2007iy,Robinson:2009xm}. 
However, the tension between the negative central charge 
and the absence of the negative norm states has not been explicitly resolved. 
Recently the three-dimensional free symplectic fermion has attracted great interest 
as a holographic dual of higher spin gravity in four-dimensional de Sitter space~\cite{Anninos:2011ui,Ng:2012xp,Das:2012dt,Anninos:2013rza,Chang:2013afa,Anninos:2014hia,Sato:2015tta,Hertog:2017ymy,Hertog:2019uhy}. 
In this dS/CFT context, the new inner product has not been fully utilized.

In this Letter we investigate the free symplectic fermion with anti-periodic boundary condition
to show that CFT$_2$ with negative central charge can be unitary. 
We review the new inner product which resolves the issue of the negative norm state
in the model~\cite{LeClair:2007iy,Robinson:2009xm}. 
We demonstrate that this new inner product follows from the connection between the path integral formalism 
and the operator formalism. 
The symplectic fermion, which is unitary with respect to the new inner product in spite of the negative central charge, 
is a counterexample of the well-known proposition that a CFT$_2$ with a negative central charge 
should have negative norm states. We clarify how the proposition can be avoided. 
We introduce $sl(2,\mathbb{R})$ invariant 
$\alpha$-vacua parametrized by 
an infinite set of real parameters $\alpha_n$ where $n$ is a positive half-integer. 
In this $\alpha$-vacuum the Hamiltonian is seemingly non-Hermitian with respect to the new inner product 
so that the energy spectrum is not necessarily real. 
We find that the energy spectrum is real in spite of the non-Hermiticity. 
We observe that the two-point function with respect to the naive norm in the $\alpha$-vacua of the symplectic fermion has divergence similar to that in the two point function of the antipodal points in the $\alpha$-vacua of de Sitter space~\cite{Goldstein:2003ut,Einhorn:2002nu,deBoer:2004nd}.

\textit{Model.}---We study the two-dimensional free symplectic fermion defined by the action
\begin{align}
	S\,=\, \int d^2x\; \partial_\mu \bar{\psi} \partial^\mu \psi\ ,
\end{align}
where $\bar{\psi}(t,\sigma)$ and $\psi(t,\sigma)$ are anti-commuting Grassmannian scalar fields~\cite{Kausch:1995py,Gaberdiel:1998ps,Kausch:2000fu,LeClair:2007iy,Robinson:2009xm}. 
In the spatial coordinate $\sigma \in S^1$, we consider the anti-boundary condition 
$\psi(t,\sigma+\ell)= - \psi (t, \sigma)$ and $\bar{\psi}(t,\sigma+\ell)= - \bar{\psi} (t, \sigma)$. 
The conjugate momenta $\Pi = {\overleftarrow{\delta} L \over \overleftarrow{\delta}\dot{\psi}}= \dot{\bar{\psi}}$ and 
$\bar{\Pi}={\overrightarrow{\delta} L \over \overrightarrow{\delta}\dot{\bar{\psi}}}=  \dot{\psi}$ 
are obtained by the right and left derivative of the Lagrangian with respect to $\dot{\psi}$ and $\dot{\bar{\psi}}$, respectively, 
for consistency with the Hermitian adjoint. 
This leads to the symplectic one-form $\pi \dot{\psi} + \dot{\bar{\psi}}\bar{\pi}$ 
in the Legendre transformation to the Hamiltonian, and the canonical anti-commutation relation reads
\begin{align}
	\{ \psi (\sigma), \Pi (\sigma')\} = i\delta(\sigma-\sigma')\;,\;\;  \{ \bar{\psi} (\sigma), \bar{\Pi} (\sigma')\} = -i \delta (\sigma-\sigma')\ . \label{eq: anticom rel}
\end{align}
Going over to Euclidean plane $z= e^{2\pi (\tau-i\sigma)\over \ell}$ with Wick rotation $\tau=it$, 
we take the mode expansion of $\psi$ and $\bar{\psi}$ as
\begin{align}
	\psi \,=\, & {i \over \sqrt{4\pi }} \sum_{n>0 } {1\over n} \big(b_n z^{-n} - c_{-n} z^n + \bar{b}_n \bar{z}^{-n} - \bar{c}_{-n} \bar{z}^n\big)\ ,\cr
	\bar{\psi} \,=\, &	{i \over \sqrt{4\pi }} \sum_{n>0 } {1\over n} \big(- b_{-n} z^n +c_n z^{-n} - \bar{b}_{-n} \bar{z}^n + \bar{c}_n \bar{z}^{-n} \big)\ ,\label{eq: mode expansion}
\end{align}
where $n$ runs over the positive half integers due to the anti-periodic boundary condition. 
From the canonical anti-commutation relation~\eqref{eq: anticom rel}, 
we can obtain the anti-commutation relation of the oscillators
\begin{alignat}{3}
	\{ b_n, b_m\}\,=\,& |n|\delta_{n+m,0}\;\;,\quad &\{ \bar{b}_n, \bar{b}_m\}\,=\,&& |n|\delta_{n+m,0}\ ,\cr
	\{ c_n, c_m\}\,=\,& -|n|\delta_{n+m,0}\;\;,\quad &\{ \bar{c}_n, \bar{c}_m\}\,=\,&& -|n|\delta_{n+m,0}\ ,\label{eq: anticomm rel}
\end{alignat}
where the negative and positive modes serve as the creation and annihilation operators, respectively. 
For more details on the quantization, please refer to Appendix A. 
Note that the anti-commutation relations of $c_n$ and $\bar{c}_n$ have a minus sign 
compared to those of $b_n $ and $\bar{b}_n$.  
This minus sign in the anti-commutation relation seemingly results in negative norm states. 
For example, using the anti-commutation relation the usual norm of the excited state $c_{-n}|0\rangle$ 
can be shown to have the sign opposite to that of the vacuum,
\begin{align}
	\langle 0 | c_{n} c_{-n} | 0 \rangle \,=\, - n\langle 0 | 0\rangle \hspace{5mm}(n>0)\ .
\end{align}
Such non-unitarity from the negative norm state has been observed 
in the higher derivative systems~\cite{Ostrogradsky:1850fid,Hawking:2001yt,Motohashi:2014opa,Woodard:2015zca,Pavsic:2016ykq,Ganz:2020skf,Lee:2021iut}. 
In such higher derivative theories, by exchanging the role of the creation and annihilation oscillators, 
one can retrieve the non-negative norm at the cost of the energy spectrum unbounded from below, 
which leads to Ostrogradsky instability~\cite{Woodard:2015zca,Ganz:2020skf}. 
However for the fermionic oscillators the negative norm cannot be cured by exchanging 
the role of the creation and the annihilation oscillators.

\textit{$\cJ$-norm and Unitarity}---The problem of the negative norm states can be resolved by 
introducing an operator $\cJ$ which is the exponentiation of the fermion number operator 
$c_n$ and $\bar{c}_n$~\cite{LeClair:2007iy,Robinson:2009xm}
\begin{align}
	\cJ \,\equiv \, e^{ \pi i \sum_n {1\over n}( c_{-n} c_n+ \bar{c}_{-n} \bar{c}_n )} \ ,
\end{align}
which is Hermitian and unitary, $\cJ^\dag =\cJ$ and $\cJ^2=1$. 
The operator $\cJ$ commutes with the oscillators $b_n$ and $\bar{b}_n$ 
while it anti-commutes with the oscillators $c_n$ and $\bar{c_n}$,
\begin{align}
	\cJ b_n \cJ\,=\, b_n\;\;,\qquad \cJ c_n \cJ\,=\, -c_n\ .
\end{align}
As in supersymmetry, one can define the $\cJ$-norm by inserting the operator $\cJ$
\begin{align}
	\langle\; \cdot \; \rangle_\cJ \,\equiv\, \langle \cJ \;\cdot \; \rangle\ .
\end{align}
Then the $\cJ$-norm of the excited states has positive $\cJ$-norm
\begin{align}
	|| \,c_{-n} |0\rangle \, ||_\cJ \,=\,\langle 0 | c_n \cJ c_{-n} |0\rangle \,=\, n ||\,  |  0 \rangle \, ||_\cJ\hspace{5mm} (n>0)\ .
\end{align}
%
Since the Hermitian adjoint follows the inner product, 
one has to define a new Hermitian adjoint $\dag_\cJ$ 
which is consistent with the new $\cJ$-norm,
\begin{align}
	\mathcal{O}^{\dag_\cJ} \,\equiv \, \cJ \cO^\dag \cJ \ .
\end{align}
For the new $\cJ$-Hermitian adjoint $\dag_\cJ$ 
we find it convenient to introduce a double-bracket notation. 
Namely, while a ket state with the double bracket is the same as the usual ket state, 
a bra state with a double bracket is defined via $\cJ$-Hermitian adjoint
\begin{align}
	|\Phi \rrangle \,\equiv\, \Phi |0\rangle  \, \quad \overset{ \dag_\cJ}{\Longrightarrow } \quad \llangle \Phi|  \,\equiv\, \langle 0 | \Phi^{\dag_\cJ} \ .
\end{align}
The inner product of double-bracket states is identical to the $\cJ$-inner product
\begin{align}
	 \llangle \Phi | \cO | \Psi \rrangle \,=\,\langle \Phi | \cO | \Psi\rangle_\cJ \ .
\end{align}
Hence double-bracket states also have the non-negative norm. Although the Hamiltonian of the symplectic fermion is Hermitian with the ordinary Hermitian adjoint, the Hermiticity of the Hamiltonian with $\cJ$-Hermitian adjoint is not straightforward in general, where more details can be found in Appendix~D. And those double-bracket bra and ket states correspond 
to the biorthogonal basis for the non-$\cJ$-Hermitian Hamiltonian~\cite{Brody_2013,Weigert:2003py}.

\textit{Connection to Path Integral}---We have introduced the $\cJ$-inner product to resolve the negative norm state problem. 
This might seem {\it ad hoc} to recover non-negative norm by modifying the theory. 
However it turns out~\cite{Lee:2021iut} that the path integral formalism of the symplectic fermion is consistent with the operator formalism with the $\cJ$-norm rather than the ordinary norm.

To understand the correspondence between the path integral and operator formalisms 
for the symplectic fermion, we define Fock states 
$|\{\nu,\mu\}\rrangle\equiv   {1\over \cN_{\{\nu,\mu\}}}\prod_{n>0} b_{-n}^{\nu_n} c_{-n}^{\mu_n} \bar{b}_{-n}^{\bar{\nu}_n}\bar{c}_{-n}^{\bar{\mu}_n} |0\rangle $ where $n$ is a positive half-integer and $\nu_n,\mu_n, \bar{\nu}_n,\bar{\mu}_n\in \{0,1\}$. We normalize the state $|\{\nu,\mu\}\rrangle$ in the double-bracket notation by choosing suitable normalization constant $\cN_{\{\nu,\mu\}}$. Note that $\llangle \{\nu,\mu\}| $ is different from $\langle \{\nu,\mu\}| $ in general. Hence the identity operator can be expressed as
\begin{align}
	\boldsymbol{I}\,=\, \sum_{\{\nu,\mu\}} |\{\nu,\mu\}\rrangle \llangle \{\nu,\mu\}| \,=\, \sum_{\{\nu,\mu\}} |\{\nu,\mu\}\rangle \langle \{\nu,\mu\}| \cJ\ .
\end{align}
In terms of ordinary bra and ket states, the operator $\cJ$ is inserted in the completeness relation, 
which makes this expression play the role of the identity operator. We also define the coherent state
\begin{align}
	|\eta, \zeta \rrangle \,\equiv\, \prod_{n>0} e^{-{1\over n} ( \eta_n b_{-n} +\zeta_n c_{-n} )} |0\rangle\ ,
\end{align}
where $\eta$ and $\zeta$ are complex numbers. 
Here we omit the anti-holomorphic part for simplicity, but one has to take it into account to connect to the path integral. 
Similarly one can define $\llangle \bar{\eta}, \bar{\zeta}|$ by using $\cJ$-Hermitian adjoint. 
In terms of the coherent state, the identity operator can be expressed as
\begin{align}
	&\boldsymbol{I}\,=\, \int \prod_{n>0} e^{ -{1\over n} (\bar{\eta}_{-n} \eta_n + \bar{\zeta}_{-n} \zeta_n )} d\eta_nd \bar{\eta}_{-n} d\zeta_n d \bar{\zeta}_{-n}    |\eta, \zeta  \rrangle \llangle\eta, \zeta |\cr
	\,=\, &\int \prod_{n>0} e^{ -{1\over n} (\bar{\eta}_{-n} \eta_n + \bar{\zeta}_{-n} \zeta_n )} d\eta_nd \bar{\eta}_{-n} d\zeta_n d \bar{\zeta}_{-n}    |\eta, \zeta  \rangle \langle\eta, \zeta | \cJ \ . \label{eq: coherent state repre of identity}
\end{align}
In the coherent state representation of the identity operator, 
the operator $\cJ$ is also inserted when we express it in terms of the ordinary coherent state.

To make contact with the path integral, 
one can insert the completeness relation~\eqref{eq: coherent state repre of identity} 
into transition amplitude $\llangle \bar{\eta}_f,\bar{\zeta}_f | \eta_i, \zeta_i\rrangle$ at each discretized time. 
The rest procedure is identical to the standard derivation of path integral 
except that the double-bracket notation, or equivalently we insert the operator $\cJ$ in the transition amplitude. 
For example, at finite temperature, one can have
\begin{align}
	\Tr\big(e^{-\beta H}\big) =\tr \big(\cJ  e^{-\beta H}\big) = \int \cD \bar{\psi} \cD \psi \cD\bar{\pi} \cD \pi  \; e^{- S [\psi,\bar{\psi},\pi,\bar{\pi} ]}\ ,
\end{align}
where the trace $\Tr$ runs over the double-bracket states while the trace $\tr$ runs over the states with single bracket. 
Note that the trace $\Tr$ corresponds to that of the biorthogonal basis. 
For more details of the derivation, see Ref.~\cite{Lee:2021iut}. 
One may express the identity operator~\eqref{eq: coherent state repre of identity} 
in terms of the ordinary coherent states $|\eta, \zeta\rangle $ and $\langle \eta, \zeta|$ 
without the $\cJ$ operator. 
However in this representation the measure becomes 
$e^{ -{1\over n} (\bar{\eta}_{-n} \eta_n - \bar{\zeta}_{-n} \zeta_n )}$. 
The asymmetry between $\eta$ and $\zeta$ in the measure makes it difficult to repeat the standard derivation.

We have seen that the $\cJ$-norm follows from the path integral of the symplectic fermion. 
Therefore the Fock states have a positive norm and positive energy, 
which implies the unitarity, at least, of the free theory.

\textit{Negative central charge and positive norm.}---Let us now discuss the Virasoro symmetry of the symplectic fermion. 
Using the anti-commutation relations~\eqref{eq: anticomm rel} of the oscillators, 
the two-point function of the primary operator $\partial \bar{\psi}$ and $\partial \psi$ of conformal dimension $1$ 
in the double-bracket notation is evaluated to yield
\begin{align}
	\llangle \partial \bar{\psi}(z) \partial \psi(w) \rrangle\,=\,   {1\over 8\pi } {\sqrt{w\over z}+ \sqrt{z\over w}\over (z-w)^2 } \ . \label{eq: two pt ftn}
\end{align}
Note that the correlation function with respect to the vacuum state with double-bracket is identical 
to that of single-bracket because the vacuum state $|0\rangle$ is invariant under the action of $\cJ$, 
{\it i.e.}, $\cJ |0\rangle =|0\rangle$.

Using the two-point function~\eqref{eq: two pt ftn}, the OPE of the energy-momentum tensor $T(z)= -4\pi : \partial \bar{\psi} \partial \psi$ is
\begin{align}
	T(z) T(w)\,\sim \,  {(-1) \over (z-w)^4} +  {2T(w)\over (z-w)^2} + {\partial T(w)\over (z-w)}\ .
\end{align}
Then we can read off the central charge $c=-2$ of the symplectic fermion. 
Note that the $\cJ$-norm is not essential in obtaining the central charge because 
the two-point function~\eqref{eq: two pt ftn} is blind to the $\cJ$ operator insertion.

Now we consider the Virasoro generator. The non-zero mode reads
\begin{align}
	L_n\,=\,  & {1\over 2} \sum_{m>0}    (b_{n-m}+c_{n-m})(b_m-c_m)   \cr
	&+ {1\over 2} \sum_{m>n}   (b_{n-m}-c_{n-m})  (b_{m}+c_{m}) \hspace{3mm} (n\ne 0)\ .
\end{align}
While the usual Hermitian adjoint of $L_n$ 
is $L_{-n}$ ({\it i.e.} $L_n^\dag=L_{-n}$), 
the $\cJ$-Hermitian adjoint of $L_n$ is different from $L_{-n}$. On the other hand, 
the zero mode $L_0$ is $\cJ$-Hermitian
\begin{align}
	L_0\,=\,  \sum_{m>0} (b_{-m} b_m - c_{-m} c_m) -{1\over 8}\ , \label{eq: generator l0}
\end{align}
where the vacuum energy density is chosen from the point-splitting regularization of 
the one-point function of the energy-momentum tensor $\llangle T(z) \rrangle = -{1\over 8 z^2}$.
%
%
Using the anti-commutation relations~\eqref{eq: anticomm rel} 
one can explicitly double-check the central charge $c=-2$ from the Virasoro algebra 
\begin{align}
	[L_n,L_m]\,=\, (n-m)L_{n+m} + {c\over 12}n(n^2-1) \delta_{n+m,0}\ .\label{eq: virasoro algebra}
\end{align}
This result seemingly contradicts the well-known proposition that CFT$_2$ with negative central charge 
has negative norm states because the symplectic fermion does not have any negative norm state 
in spite of the negative central charge. 
We find that the standard proof for the proposition has a "loophole" for the case of the symplectic fermion.

The standard proof considers the norm of the state $L_{-n} |h\rangle$ ($n>0$) where $|h\rangle$ 
is a primary state with conformal dimension $h$. 
Using the Virasoro algebra~\eqref{eq: virasoro algebra} 
one can show that the norm of the state $L_{-n}|h\rangle$ has opposite sign to that of $|h\rangle$ for sufficiently large $n$. 
%
%
However for the symplectic fermion one has to use $\cJ$-Hermitian adjoint as well as $\cJ$-norm, or equivalently, 
double-bracket states. Therefore the correct norm of the state $L_{-n}|h\rangle$ is
\begin{align}
	\llangle L_{-n} h | L_{-n}h \rrangle \,=\, \llangle h | L_{-n}^{\dag_\cJ} L_{-n} |h \rrangle\,=\, \llangle h | \cJ L_n \cJ L_{-n} |h\rrangle\ .
\end{align}
Since $L_n^{\dag_\cJ}\ne L_{-n}$ for $n\ne 0$ as we observed above, 
one cannot use the Virasoro algebra to get the proposition. 
Therefore we conclude that the symplectic fermion is a "counterexample" for the non-unitarity of CFT$_2$ 
with negative central charge.

The unitarity of the symplectic fermion implies that physical quantities will be well-defined.
For example, the entanglement entropy should be positive in unitary theory. 
For the case of negative central charge the entanglement entropy of a subsystem of length $a$ is not proportional to the central charge; but it is given by~\cite{Bianchini:2014uta,Bianchini:2015uea,Couvreur:2016mbr}
\begin{align}
	S_{{\rm EE}}(a)\,=\, {c_{\text{\tiny eff}}\over 3} \log \bigg({ a \over  \epsilon } \bigg)\ ,
\end{align}
where $\epsilon$ is the ultraviolet cut-off. The effective central charge denoted by $c_{\text{\tiny eff}}$ is defined by $c_{\text{\tiny eff}}= c- 24\Delta_{\text{\tiny min}}$ where $\Delta_{\text{\tiny min}}$ is the lowest holomorphic conformal dimension. For the case of the symplectic fermion the vacuum energy density in Eq.~\eqref{eq: generator l0} implies that the identity operator has the lowest conformal dimension $\Delta_{\text{\tiny min}}=-1/8$, and we have $c_{\text{\tiny eff}}=1$. Thus the entanglement entropy is positive as expected for a unitary theory. 
However the positive effective central charge does not always imply the unitarity. 
For instance, the Lee-Yang model with $c=-22/5$ has positive effective $c_{\text{\tiny eff}}=2/5$ in spite of the non-unitary.

\textit{$\alpha$-vacua}---The $\cJ$-Hermiticity of $L_0$ and the unitarity is non-trivial in the alternative mode expansion~\eqref{eq: mode expansion} of the $\psi$ and $\bar{\psi}$. To see this issue, we define Bogoliubov generator $\cG_\alpha$ by
\begin{align}
	\cG_\alpha\,\equiv\,  i \sum_{n>0} {\alpha_n \over n} \big( b_{-n}c_{-n} + b_n c_n  \big)\ ,
\end{align}
where  $\alpha_n\in \mathbb{R}$ ($n=1/2,3/2, \cdots$). Note that $\cG_\alpha$ is Hermitian but not $\cJ$-Hermitian. The adjoint action of $\cJ$ on $\cG_\alpha$ flips the sign of all $\alpha$'s, {\it i.e.}, $\cJ \cG_\alpha \cJ= \cG_{-\alpha}$. Similarly we can define $\bar{\cG}_\alpha$ for the anti-holomorphic oscillators, 
but we omit the anti-holomorphic contributions for simplicity which is parallel to the holomorphic calculation. Since $\cG_\alpha$ generates the canonical (Bogoliubov) transformation, one may take the mode expansion of $\psi$ and $\bar{\psi}$ in terms of $\tilde{b}_n\equiv e^{-i\cG_\alpha }b_n e^{i\cG_\alpha}$ and $\tilde{c}_n\equiv e^{-i\cG_\alpha }c_n e^{i\cG_\alpha}$ instead of $b_n$ and $c_n$. 
%
%
In this new mode expansion, $L_0$ is expressed as
\begin{align}
	L_0 = \sum_{n>0} \big[ &\cosh 2\alpha_n (\tilde{b}_{-n} \tilde{b}_n - \tilde{c}_{-n}\tilde{c}_n) \cr
	&+ \sinh 2\alpha_n (\tilde{b}_{-n}\tilde{c}_{-n} + \tilde{c}_n \tilde{b}_n) \big]\ , \label{eq: transformed l0}
\end{align}
up to vacuum energy density constant. 
The oscillators $\tilde{b}_n$ and $\tilde{c}_n$ also depend 
on $\alpha$'s, 
and the adjoint action of $\cJ$ on $\tilde{b}_n$ and $\tilde{c}_n$ flips 
the sign of $\alpha_n$'s, {\it i.e.}, 
$\cJ \tilde{b}_n^{(\alpha)}\cJ = \tilde{b}_n^{(-\alpha)}$ 
and $\cJ \tilde{c}_n^{(\alpha)}\cJ =- \tilde{c}_n^{(-\alpha)}$. 
Hence it is more convenient to define a new operator 
$\tilde{\cJ}= \exp (\pi i \sum_n {1\over n} \tilde{c}^{(\alpha)}_{-n}\tilde{c}^{(\alpha)}_n)$ 
instead of the original $\cJ$ operator. 

Repeating the same procedure with $\tilde{\cJ}$-inner product and $\tilde{\cJ}$-Hermitian 
adjoint, we note that $L_0$ is not $\tilde \cJ$-Hermitian anymore though $L_0$ is still Hermitian. Therefore the eigenvalue of $L_0$ is not necessarily real. The non-$\tilde\cJ$-Hermiticity of $L_0$ can arise because the Bogoliubov transformation generated by $\cG_\alpha$ is not $\cJ$-unitary transformation but a similarity transformation.

Using the bra and ket states with double-bracket, 
the matrix elements of $L_0$ can be evaluated, and its eigenvalues are identical to the original real eigenvalues from the $b_n$ and $c_n$ oscillators,
where more details can be found in Appendix~D. 
A similar phenomenon has been observed in~\cite{Lee:2021iut} 
with a quantum mechanical model 
where a non-$\cJ$-Hermitian Hamiltonian 
can have real eigenvalues 
when there exists a Bogoliubov transformation 
that makes the Hamiltonian $\cJ$-Hermitian. 
And if such a Bogoliubov transformation does not exist, 
the Hamiltonian develops complex energy spectrum. 
Since the expression \eqref{eq: transformed l0} is obtained by the Bogoliubov transformation of the $\cJ$-Hermitian operator, it is not surprising to have real eigenvalues which are identical to the original ones.

Going back to the original $\cJ$ operator 
and the corresponding $\cJ$-norm, 
we consider the $\alpha$-vacuum $|\alpha\rrangle$ which is annihilated by $\tilde{b}_{n}$ and $\tilde{c}_{n}$ for $n>0$. The $\alpha$-vacuum is related to the vacuum $|0\rangle$ by the Bogoliubov transformation
\begin{align}
	|\alpha\rrangle =  {e^{-i \cG_\alpha }\over \sqrt{\cN} }|0\rangle = \prod_{n>0} \bigg( {\cosh \alpha_n  + \sinh\alpha_n {1\over n} b_{-n} c_{-n} \over \sqrt{\cosh 2\alpha_n}} \bigg) \; |0\rangle \ , \label{eq: alpha vacuum}
\end{align}
where $\cN= \prod_{n>0} \cosh 2\alpha_n$ is the normalization constant. Under the assumption that the vacuum $|0\rangle$ is invariant under the under the action of $\cJ$, the $\alpha$-vacuum is not, {\it i.e.} $\cJ|\alpha\rrangle = |-\alpha\rrangle$. 

The operators 
$J^0=\sum_{n>0} {1\over 2n} (b_{-n}b_n+c_{-n}c_n)$ 
and $J^\pm=\pm \sum_{n>0} {1\over n} b_{\mp n} c_{\pm n}$ 
form $sl(2,\mathbb{R})$ algebra, and the vacuum $|0\rangle$ is $sl(2,\mathbb{R})$ invariant ground state. Since $\cG_\alpha$  commutes with the $sl(2,\mathbb{R})$ generators, the $\alpha$-vacuum is also the $sl(2,\mathbb{R})$ invariant ground state.

Note that the $\alpha$-vacuum is 
the maximally entangled state of 
the Fock spaces $\cH_{b}$ and $\cH_{c}$ 
created by the oscillators $b$ and $c$, respectively. 
The usual maximally entangled states obtained 
by Bogoliubov transformations do not need 
the non-trivial normalization constant $\cN$, 
{\it i.e.}, $\cN=1$. 
However, the non-trivial normalization constant 
is necessary for the symplectic fermion 
because the Bogoliubov transformation in our case 
is not a unitary but similarity one. 
By tracing out the $\cH_c$, 
the reduced density matrix $\rho_b$ 
of the pure state $|\alpha\rrangle$ reads
\begin{align}
	\rho_b\,=\, \bigotimes_{n>0}\bigg( {\cosh^2 \alpha_n \over \cosh 2\alpha_n} |0\rangle \langle 0 | + {\sinh^2 \alpha_n \over \cosh 2\alpha_n } b_{-n}|0\rangle \langle 0| b_n \bigg)\ .
\end{align}
If we choose a specific value of $\alpha_n= e^{-{\beta| n | \over 2}}$, the reduced density matrix is identical to the thermal density matrix of the fermi oscillator $b$ with temperature $\beta^{-1}$, and the $\alpha$-vacuum corresponds to the thermofield dynamics state~\cite{Takahashi:1996zn}.

We now evaluate the two-point function of the primary operator with respect to the $\alpha$-vacuum. We find that
\begin{align}
	&\llangle \partial \bar{\psi}(z)  \partial \psi (w)  \rrangle_{\alpha} \cr
	=&  {1\over 4\pi zw} \sum_{n>0} n \bigg[ \bigg({w\over z}\bigg)^n {\cosh^2\alpha_n\over \cosh 2\alpha_n} - \bigg({z\over w}\bigg)^n {\sinh^2\alpha_n\over \cosh 2\alpha_n} \bigg]\ .
\end{align}
If all $\alpha_n$ are set to the same value $\alpha$, 
the two-point function is independent of $\alpha$ 
and it reproduces the two-point function~\eqref{eq: two pt ftn} 
with respect to the vacuum state. 
On the other hand, one can also evaluate the two-point function 
with the naive norm without $\cJ$ insertion 
which is not consistent with the path integral formulation 
and leads to negative norm states. 
With the naive norm, 
we have the ordinary Hermitian adjoint, 
and the ket state of $\alpha$-vacuum 
does not need the non-trivial normalization 
constant $\cN$ as usual.  
Using the $\alpha$-vacuum~\eqref{eq: alpha vacuum} 
with $\cN=1$, we obtain
\begin{align}
	&\langle \partial \bar{\psi}(z)  \partial \psi (w)  \rangle_\alpha \cr
	=& {1\over 4\pi z w}\sum_{n>0} n \bigg[ \bigg({w\over z}\bigg)^n\cosh^2\alpha_n+ \bigg({z\over w}\bigg)^n\sinh^2\alpha_n    \cr
	&\hspace{15mm}+ \big((zw)^n + (zw)^{-n} \big)\sinh \alpha_n \cosh\alpha_n   \bigg]\ .
\end{align}
This result with naive norm contains the power series in $zw$ in addition to that in $z/w$. This implies that the correlation function could diverge at $z=1/w$ as well as $z=w$. To see this divergence explicitly, we set all $\alpha_n$ parameters to be the same value $\alpha$, and we obtain
\begin{align}
	\langle \partial\bar{\psi}(z)  \partial \psi (w)  &\rangle_{\alpha_n=\alpha} = {1\over 8\pi }  { \sqrt{z\over w}+ \sqrt{w\over z} \over (z-w)^2} \cosh(2\alpha) \cr
	&+ {1\over 8\pi } {1\over w^2} { \sqrt{z w}+ \sqrt{1\over z w} \over \big(z-{1\over w}\big)^2} \sinh(2\alpha)\ .
\end{align}

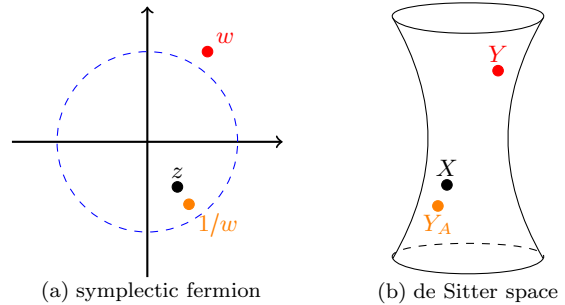
\begin{figure}[t!]
\centering

\subfigure[\;symplectic fermion]{
\begin{minipage}[c]{0.45\linewidth}
	 \begin{tikzpicture}

	 \def\tikzr{1.2}
	\def\tikzl{1.8}
	\def\tikzrho{0.07}
	
	\def\tikzyx{0.8}
	\def\tikzyy{1.2}
	
	\def\tikzyinvx{0.553846}
	\def\tikzyinvy{-0.830769}
	
	\def\tikzxx{0.4}
	\def\tikzxy{-0.6}
	
	\draw[thick,->,black] (-\tikzl,0)--(\tikzl,0) node[below] {}; 
    \draw[thick,->,black] (0,-\tikzl)--(0,\tikzl) node[left] {}; 
    \draw[blue,dashed] (0,0) circle (\tikzr);
    \draw [fill,red] (\tikzyx,\tikzyy) circle [radius=\tikzrho] node[above right] {$w$};
    \draw [fill,orange] (\tikzyinvx,\tikzyinvy) circle [radius=\tikzrho] node[below right] {$1/w$};
    \draw [fill] (\tikzxx,\tikzxy) circle [radius=\tikzrho] node[above] {$z$};
    \end{tikzpicture}

\end{minipage}
}
\subfigure[\;de Sitter space]{
\begin{minipage}[c]{0.45\linewidth}
	 \begin{tikzpicture}
	\def\tikzlx{1.0}
	\def\tikzly{1.6}
	
	\def\tikzrho{0.07}

	\def\tikzyx{0.4}
	\def\tikzyy{0.9}
	
	\def\tikzyinvx{-0.4}
	\def\tikzyinvy{-0.9}
	
	\def\tikzxx{-0.28}
	\def\tikzxy{-0.62}
	
	\draw (\tikzlx,-\tikzly) to[out=120,in=-120] (\tikzlx,\tikzly);
	\draw (-\tikzlx,-\tikzly) to[out=60,in=-60] (-\tikzlx,\tikzly);
	
	\draw (-\tikzlx,\tikzly) to[out=120,in=60, looseness=0.4] (\tikzlx,\tikzly);
	\draw (-\tikzlx,\tikzly) to[out=-60,in=-120, looseness=0.4] (\tikzlx,\tikzly);
	
	\draw [dashed] (-\tikzlx,-\tikzly) to[out=120,in=60, looseness=0.4] (\tikzlx,-\tikzly);
	\draw (-\tikzlx,-\tikzly) to[out=-60,in=-120, looseness=0.4] (\tikzlx,-\tikzly);
	
    \draw [fill,red] (\tikzyx,\tikzyy) circle [radius=\tikzrho] node[above ] {$Y$};
    \draw [fill,orange] (\tikzyinvx,\tikzyinvy) circle [radius=\tikzrho] node[below] {$Y_A$};
    \draw [fill] (\tikzxx,\tikzxy) circle [radius=\tikzrho] node[above] {$X$};
	
	\end{tikzpicture}
\end{minipage}
\label{fig:1b}}
	\caption{Two point function in the $\alpha$-vacuum of symplectic fermion and scalar field in the $\alpha$-vacuum of de Sitter space. (a) With the naive inner product, the two point function $\langle \partial\bar{\psi}(z)  \partial \psi (w)  \rangle_{\alpha_n=\alpha}$ with respect to $\alpha$-vacuum diverges as $z$ approaches to $1/w$. (b) Two point function of scalar field in the $\alpha$-vacuum of de Sitter space also diverges when $X$ is close to the antipodal point $Y_A$ of $Y$.}
	\label{fig:1}
\end{figure}

The divergence of the naive two-point function 
at $zw=1$ has been observed in the two-point function 
of the free scalar field in the $\alpha$-vacua 
of de Sitter space~\cite{Goldstein:2003ut,Einhorn:2002nu,deBoer:2004nd}, which is demonstrated in in Fig.~\ref{fig:1}. 
 The two-point function with respect to the $\alpha$-vacuum 
of the de Sitter space diverges 
when one point is identical or 
antipodal to the other in the de Sitter space, where more detailed comparison can be found in Appendix C. 
The Bunch-Davies vacuum~\cite{Bunch:1978yq} 
is free of the divergence of the antipodal points, 
which is analogous to the vacuum $|\alpha=0\rangle=|0\rangle$ 
in the symplectic fermion with naive norm.

\textit{Discussion.}---In this Letter, we have explained that the symplectic fermion is a unitary CFT$_2$ in spite of the negative central charge $c=-2$. The $\cJ$-norm following from the path integral formulation makes the theory unitary, and the corresponding $\cJ$-Hermitian adjoint plays a key role in avoiding the well-known proposition on the existence of negative norm states in the CFT$_2$ with negative central charge. We have also analyzed the $sl(2,\mathbb{R})$ invariant $\alpha$-vacua in which non-$\cJ$-Hermitian Hamiltonian can retrieve the real energy spectrum. We have compared
 the two-point function with the naive norm to that of de Sitter space.

Our work suggests a new class of unitary CFT with negative central charge. It might be possible to explore a lattice model or an interacting continuum CFT$_2$ which is unitary but has negative central charge. The divergence of two point function at the antipodal points in the $\alpha$-vacua of de Sitter space is still an open problem. In the symplectic fermion, the analogous problem is cured by the $\cJ$-inner product. Thus it is highly interesting to find an alternative inner product in the de Sitter space which might shed light on revisiting the $\alpha$-vacua issue in the de Sitter space.

We thank Changrim Ahn for discussion. 
S.R. is supported by
the National Science Foundation under award number
DMR-2001181, and by a Simons Investigator Grant from
the Simons Foundation (Award Number: 566116).
J.Y. was supported by the National Research Foundation of Korea (NRF) 
grant funded by the Korean government (MSIT) 
(No.\ 2019R1F1A1045971, 2022R1A2C1003182). This research was supported in part by the International Centre for Theoretical Sciences (ICTS) for the program "Nonperturbative and Numerical Approaches to Quantum Gravity, String Theory and Holography " (code: ICTS/numstrings-2022/8).
J.Y. is supported by an appointment to the JRG Program 
at the APCTP through the Science and Technology 
Promotion Fund and Lottery Fund of the Korean Government. 
This is also supported by the Korean Local Governments 
- Gyeongsangbuk-do Province and Pohang City. This work is also supported by Korea Institute for Advanced Study (KIAS)
 grant funded by the Korean government.

\begin{widetext}

\appendix

\section{Appendix A: Quantization of Two-dimensional Symplectic Fermion}
\label{App:quantization}

We review the quantization of the two-dimensional free symplectic fermion given by
\begin{align}
	S\,=\,\int d^2x\; \big( \partial^\mu \bar{\psi}\partial_\mu \psi \big)\,=\, \int d^2x\; \big( \dot{\bar{\psi}} \dot{\psi} - \bar{\psi}'\psi'\big)\ .
\end{align}
where $\dot{\psi}\equiv \partial_t \psi$ and $\psi'\equiv \partial_\sigma \psi$ etc. Let us now consider the mode expansion with respect to the spatial coordinate $\sigma \in S^1$ with period $\ell$
\begin{align}
	\psi(t,\sigma)\,=\, \sum_{n\in\mathbb{Z}+{1\over 2}}  \psi_{n}(t) e^{ 2\pi i n \sigma\over \ell}\;\;,\quad \bar{\psi}(t,x)\,=\, \sum_{n\in\mathbb{Z}+{1\over 2}} \bar{\psi}_{n}(t) e^{ 2\pi i n \sigma\over \ell} \ ,
\end{align}
where we consider anti-periodic boundary condition so that $n$ is a half-integer. In terms of the Fourier modes, the action reads
\begin{align}
	S\,=\, \ell \int dt \sum_{n\in \mathbb{Z}+{1\over 2}} \bigg[ \,\dot{\bar{\psi}}_{n} \dot{\psi}_{-n} - \bigg({2\pi n \over \ell}\bigg)^2\bar{\psi}_{n}\psi_{-n}\, \bigg]\ .
\end{align}
One can obtain the conjugate momentum $\Pi$ and $\bar{\Pi}$ by the right and left derivative of the Lagrangian $L$ with respect to $\dot{\psi}_n$ and $\dot{\bar{\psi}}_n$, respectively
\begin{align}
	\Pi_n\,=\, {\overleftarrow{\delta} L \over \overleftarrow{\delta}\dot{\psi}_n}\,=\, \ell\, \dot{\bar{\psi}}_{-n}\;\;,\quad\bar{\Pi}_n\,=\, {\overrightarrow{\delta}L \over \overrightarrow{\delta}\dot{\bar{\psi}}_n}\,=\, \ell\, \dot{\psi}_{-n}\ ,
\end{align}
and the canonical anti-commutation relation is given by
\begin{align}
	\{ \psi_n, \Pi_m\}\,=\, i\,\delta_{n,m}\;\;,\qquad \{ \bar{\psi}_n,\bar{\Pi}_n\}\,=\, - i \, \delta_{n,m}\ .
\end{align}
Note that we have the minus sign on the right-hand side of the anti-commutation relation of $\bar{\psi}_n$ and $\bar{\Pi}_n$ which makes Hermitian adjoint of the anti-commutation relation consistent. Accordingly, the consistent symplectic one-form in the Legendre transformation to the Hamiltonian is $\Pi_n \dot{\psi}_n+ \dot{\bar{\psi}}_n \bar{\Pi}_n$, and the Hamiltonian reads
\begin{align}
	H\,=\,\sum_{n\in \mathbb{Z}+{1\over 2} } \big(\Pi_{-n} \dot{\psi}_{-n}+ \dot{\bar{\psi}}_n \bar{\Pi}_n\big) - \mathcal{L}\,=\,{1\over \ell }\sum_{n\in \mathbb{Z}+{1\over 2} } \big(\, \Pi_{-n}\bar{\Pi}_{n} + (2\pi n)^2 \bar{\psi}_{n}\psi_{-n}\,\big)\ .
\end{align}
Let us now define oscillators $\tilde{b}_n\, ,\; \tilde{b}_n^\dag\, ,\; \tilde{c}_n$ and $\tilde{c}_n^\dag$ by
\begin{align}
	\tilde{b}_n\,=\, &{1\over \sqrt{4\pi |n|}}\big( 2\pi |n| \psi_{n} + i \bar{\Pi}_{-n} \big)\;\;,\quad \tilde{c}_n\,=\, {1\over \sqrt{4\pi |n|}}\big( 2\pi |n| \bar{\psi}_{n} + i\Pi_{-n} \big)\ ,\\
	\tilde{b}_n^\dag\,=\, &{1\over \sqrt{4\pi |n|}}\big( 2\pi |n| \bar{\psi}_{-n} - i \Pi_{n} \big)\;\;,\quad \tilde{c}_n^\dag\,=\, {1\over \sqrt{4\pi |n|}}\big( 2\pi |n| \psi_{-n} - i \bar{\Pi}_{n} \big)\ ,
\end{align}
where their anti-commutation relation reads
\begin{align}
	\{\tilde{b}_n, \tilde{b}_m^\dag\} \,=\, \delta_{n,m}\;\;,\quad \{\tilde{c}_n, \tilde{c}_m^\dag\} \,=\, -\delta_{n,m}\ .
\end{align}
For the usual convention in CFT$_2$, we redefine oscillators $b_n\,,\; c_n$ and $\bar{b}_n\,,\; \bar{c}_n$ as follows.
\begin{align}
	b_n\,=\,& \begin{cases}
	\;\; -i \sqrt{n} \;\tilde{b}_n &\quad (n>0)\\
	\;\; i\sqrt{-n} \;\tilde{b}_{-n}^\dag &\quad (n<0)\\
	\end{cases}\quad,\qquad \bar{b}_n\,=\, \begin{cases}
	\;\;-i \sqrt{n} \;\tilde{b}_{-n} &\quad (n>0)\\
	\;\; i\sqrt{-n} \;\tilde{b}_{n}^\dag &\quad (n<0)\\
	\end{cases}\ ,\\
	c_n\,=\,& \begin{cases}
	\;\; -i \sqrt{n} \;\tilde{c}_n &\quad (n>0)\\
	\;\; i\sqrt{-n} \;\tilde{c}_{-n}^\dag &\quad (n<0)\\
	\end{cases}\quad,\qquad	\bar{c}_n\,=\, \begin{cases}
	\;\;-i \sqrt{n} \;\tilde{c}_{-n} &\quad (n>0)\\
	\;\; i\sqrt{-n} \;\tilde{c}_{n}^\dag &\quad (n<0)\\
	\end{cases}\ ,
\end{align}
and their anti-commutation relations become
\begin{alignat}{3}
	\{ b_n, b_m\}\,=\,& |n|\delta_{n+m,0}\;\;,\quad &&\{ \bar{b}_n, \bar{b}_m\}\,=\, |n|\delta_{n+m,0}\ , \label{eq: anti comm osc 1}\\
	\{ c_n, c_m\}\,=\,& -|n|\delta_{n+m,0}\;\;,\quad &&\{ \bar{c}_n, \bar{c}_m\}\,=\, -|n|\delta_{n+m,0} \ . \label{eq: anti comm osc 2}
\end{alignat}
Note that the right-hand side of the anti-commutation relation of $c_n$ and $\bar{c}_n$  has the opposite sign compared to the usual fermi oscillator. In terms of the oscillators, the Hamiltonian is written as
\begin{align}
	H\,=\, {2\pi \over \ell} \sum_{n>0} (b_{-n}b_n +\bar{b}_{-n} \bar{b}_n - c_{-n}c_n - \bar{c}_{-n} \bar{c}_n)\ .
\end{align}
Note that the Hamiltonian is $\cJ$-Hermitian. One can see that $b_{-n}, \bar{b}_{-n}, c_{-n}$ and $\bar{c}_{-n}$ is a creation operator for $n>0$. 
\begin{align}
	[H,b_{-n}]\,=\, & {2\pi \over \ell}| n| b_{-n}\;\;,\quad [H,c_{-n}]\,=\,   {2\pi \over \ell} |n| c_{-n}\quad \mbox{\it etc.}\ .
\end{align}
One can express $\psi$ and $\bar{\psi}$ in terms of the oscillators as
\begin{align}
	\psi \,=\, &{i\over \sqrt{4\pi}} \sum_{n>0} {1\over n} \big( b_n - \bar{c}_{-n} \big)\; e^{2\pi i n x \over \ell}+ {i\over \sqrt{4\pi}} \sum_{n<0} {1\over n} \big(-\bar{b}_{-n} + c_n \big)\; e^{2\pi i n x \over \ell} \ ,\\
	\bar{\psi}\,=\, & {i\over \sqrt{4\pi}} \sum_{n>0} {1\over n} \big( -\bar{b}_{-n} + c_n  \big)\; e^{2\pi i n \sigma \over \ell} +  {i\over \sqrt{4\pi}} \sum_{n<0} {1\over n} \big( b_n  - \bar{c}_{-n} \big)\; e^{2\pi i n \sigma \over \ell} \ .
\end{align}
Using the time evolution of the oscillators, we have
\begin{align}
	\psi(t,\sigma) \,=\, &{i\over \sqrt{4\pi}} \sum_{n>0} {1\over n} \bigg( b_n e^{ {2\pi i n\over \ell} (-t + \sigma) }  - c_{-n} e^{- {2\pi i n\over \ell} (-t+ \sigma) } + \bar{b}_{n}e^{- {2\pi i n\over \ell} ( t+ \sigma  ) } - \bar{c}_{-n} e^{ {2\pi i n\over \ell} (t+ \sigma )} \bigg)\ ,\\
	\bar{\psi}(t,\sigma) \,=\, & {i\over \sqrt{4\pi}} \sum_{n>0} {1\over n} \bigg(   -b_{-n} e^{-{2\pi i n\over \ell} (-t+\sigma)} + c_ne^{ {2\pi i n\over \ell} (-t+\sigma) }  -\bar{b}_{-n}e^{ {2\pi i n\over \ell} ( t+\sigma ) }   + \bar{c}_{n} e^{- {2\pi i n\over \ell}  (t+\sigma) }\bigg) \ .
\end{align}

Let us now proceed to Euclidean space by the Wick rotation $t=-i\tau$. By transforming to complex plane $z$ and $\bar{z}$ defined by
\begin{align}
	z\,\equiv\, e^{2\pi (\tau-ix)\over \ell}\quad,\qquad 	\bar{z}\,\equiv \, e^{2\pi (\tau+ ix)\over \ell} \ ,
\end{align}
the mode expansion of $\psi$ and $\bar{\psi}$ reads
\begin{align}
	\psi (z,\bar{z}) \,=\, & {i \over \sqrt{4\pi }} \sum_{n>0 } {1\over n} \big(b_n z^{-n} - c_{-n} z^n + \bar{b}_n \bar{z}^{-n} - \bar{c}_{-n} \bar{z}^n\big)\ , \cr
	\bar{\psi} (z,\bar{z}) \,=\, &	{i \over \sqrt{4\pi }} \sum_{n>0 } {1\over n} \big(- b_{-n} z^n+ c_n z^{-n} - \bar{b}_{-n} + \bar{c}_n \bar{z}^{-n}  \bar{z}^n\big)\ .\label{eq: mode expansion of psi}
\end{align}

Using the anti-commutation relations~\eqref{eq: anti comm osc 1}$\sim$\eqref{eq: anti comm osc 2} of the oscillators, one can evaluate the two-point function with the $\cJ$-norm, or equivalently with the double-bracket notation. For example, the two point function $\llangle \partial \psi(z) \psi(w) \rrangle$ vanishes as expected
\begin{align}
	\llangle \partial \psi(z) \psi(w) \rrangle\,=\, {1\over 4\pi } \sum_{n,m>0} {1\over m} \llangle (b_n z^{-n-1} +c_{-n} z^{n-1})(b_mw^{-m}-c_{-m}w^{m})\rrangle \,=\,0\ .
\end{align}
The non-vanishing two-point functions are evaluated to yield 
\begin{align}
	\llangle \partial \bar{\psi}(z) \psi(w) \rrangle\,=\, &{1\over 4\pi } \sum_{n,m>0} {1\over m} \llangle (b_{-n}z^{n-1}+c_nz^{-n-1})(b_mw^{-m}-c_{-m}w^{m})\rrangle \,=\,   {1\over 4\pi } \sum_{n={1\over 2} }^\infty {w^n \over z^{n+1}} \,=\,  {1\over 4\pi } {\sqrt{w\over z}\over z-w }\ ,
\end{align}
and
\begin{align}
	\llangle \partial \bar{\psi} (z) \partial \psi (w) \rrangle\,=\, &-{1\over 4\pi } \sum_{n,m>0} \llangle (b_{-n}z^{n-1}+c_nz^{-n-1})(b_mw^{-m-1}+c_{-m}w^{m-1})\rrangle \,=\,  {1\over 4\pi } \sum_{n={1\over 2} }^\infty n { w^{n-1} \over z^{n+1} } \,=\, {1\over 8\pi } {\sqrt{w\over z}+ \sqrt{z\over w}\over (z-w)^2 }  \ . \label{eq: two ptn ftn} 
\end{align}

\section{Appendix B: Virasoro Symmetry}
\label{App:virasoro}

In this section, we discuss the Virasoro symmetry of the symplectic fermion. From the Euclidean action on the complex plane $(z,\bar{z})$
\begin{align}
	S\,=\,\int dzd\bar{z} \; \big(2 \partial \bar{\psi} \bar{\partial}\psi + 2  \bar{\partial} \bar{\psi} \partial\psi \big)\ ,
\end{align}
the energy momentum tensor reads 
\begin{align}
	T^{\bar{z}\bar{z}}\,=\,& 2{\overleftarrow{\delta} \cL \over \overleftarrow{\delta} \bar{\partial} \psi} \partial \psi + 2{\overleftarrow{\delta} \cL \over \overleftarrow{\delta} \bar{\partial} \bar{\psi}} \partial \bar{\psi} \,=\, 8  \partial \bar{\psi} \partial \psi \ , \\
	T^{zz}\,=\,&2{\overleftarrow{\delta} \cL \over \overleftarrow{\delta} \partial \psi}\bar{\partial} \psi + 2{\overleftarrow{\delta} \cL \over \overleftarrow{\delta} \partial \bar{\psi}} \bar{\partial} \bar{\psi} \,=\,8  \bar{\partial} \bar{\psi} \bar{\partial} \psi \ ,\\
	T^{z\bar{z}}\,=\,& 2{\overleftarrow{\delta} \cL \over \overleftarrow{\delta} \partial \psi} \partial \psi + 2{\overleftarrow{\delta} \cL \over \overleftarrow{\delta} \partial \bar{\psi}} \partial \bar{\psi}- 2\cL\,=\,0\ .
\end{align}
The holomorphic energy momentum tensor is defined by 
\begin{align}
	T(z)\,\equiv\, - 2\pi T_{zz}=-{\pi \over 2} T^{\bar{z}\bar{z}}\,=\, - 4\pi  : \partial \bar{\psi} \partial \psi:\ .
\end{align}
Using Eq.~\eqref{eq: two ptn ftn}, one can obtain the OPE of $T(z)$ and the $\partial \psi(w)$,
\begin{align}
	T(z)\partial \psi(w)\,=\,& - 4\pi  : \partial \bar{\psi}(z) \partial \psi (z): \partial\psi(w) \,\sim\, {1\over (z-w)^2 }\partial \psi (z)\,\sim\, { \partial \psi(w)\over (z-w)^2} + { \partial^2 \psi(w) \over z-w}\ .
\end{align}
From the OPE we confirm that $\partial \psi$ is a primary operator of dimension $h=1$. Furthermore, the OPE of two energy-momentum tensors is 
\begin{align}
	T(z) T(w)\,=\,& 16\pi^2 : \partial \bar{\psi}(z) \partial \psi(z): : \partial \bar{\psi}(w) \partial \psi(w):\,\sim \, -{ 1 \over (z-w)^4}  + {4\pi  \over (z-w)^2} \big(-:\partial\bar{\psi}(z)\partial \psi(w):+ :\partial\psi(z)\partial \bar{\psi}(w):\big)\ ,\cr
	\,\sim \, & -{ 1 \over (z-w)^4}  - {8\pi  :\partial\bar{\psi} (w)\partial \psi (w): \over (z-w)^2} - {4\pi  \over z-w} \big(:\partial\bar{\psi}(w)\partial^2 \psi(w):+ :\partial^2\bar{\psi}(w)\partial \psi(w):\big)\ ,\cr
	\,=\,& {(-1) \over (z-w)^4} +  {2T(w)\over (z-w)^2} + {\partial T(w)\over (z-w)}\ .
\end{align}
This shows that the central charge of the symplectic fermion is 
\begin{align}
	c \,=\, -2\ .
\end{align}
From the mode expansion of the normal-ordered energy-momentum tensor 
\begin{align}
	T(z)\,=\, & -4\pi  : \partial\bar{\psi}\partial \psi: \,=\, \sum_{n,m>0} :  (b_{-n} z^{n-1}+c_n z^{-n-1})(b_m z^{-m-1}+c_{-m} z^{m-1}) : \ ,
\end{align}
one can read off the Virasoro generator $L_n$. We find
\begin{align}
	L_n \,=\,  {1\over 2} \sum_{m>0}   :  (b_{n-m}+c_{n-m})(b_m-c_m) : - {1\over 2} \sum_{m<0}  :  (b_{n-m}+c_{n-m})(b_m-c_m) :\ .
\end{align}
For $n\ne 0$,  we have
\begin{align}
	L_n\,=\,  {1\over 2} \sum_{m>0}    (b_{n-m}+c_{n-m})(b_m-c_m)   + {1\over 2} \sum_{m>n}   (b_{n-m}-c_{n-m})  (b_{m}+c_{m})\ , \label{eq: nonzero virasoro mode}
\end{align}
and for $n=0$, we have
\begin{align}
	L_0\,=\, & {1\over 2} \sum_{m} {m \over |m| } :  (b_{-m}+c_{-m})(b_m-c_m) : \,=\,  {1\over 2} \sum_{m>0}   : (b_{-m}+c_{-m})(b_m-c_m)  : + {1\over 2} \sum_{m>0}  : (b_{-m}-c_{-m})  (b_{m}+c_{m}): \ , \cr
	\,=\, & \sum_{m>0} (b_{-m} b_m - c_{-m} c_m) -{1\over 8}\ .
\end{align}
Here the vacuum energy density $-{1\over 8}$ is determined by the one-point function of the energy momentum tensor with the point-splitting regularization 
\begin{align}
	\langle T(z) \rangle \,=\, \lim_{\epsilon\to 0} \bigg( -4\pi  \langle \partial \bar{\psi}(z+\epsilon)\partial\psi(z) \rangle +{1\over \epsilon^2} \bigg) \,=\, -{1\over 8 z^2} \ .\label{eq: one pt ftn T}
\end{align}
Note that the non-zero mode $L_n$ ($n\ne 0$) \eqref{eq: nonzero virasoro mode} is non-$\cJ$-Hermitian because of the terms which are linear in $c_m$'s. As expected, the Virasoro generators satisfy the Virasoro algebra with $c=-2$
\begin{align}
	[L_n,L_m]\,=\,  (n-m)L_{n+m} +{c\over 12} n(n^2-1)\delta_{n+m,0} \ ,
\end{align}
To confirm it explicitly, we confirm
\begin{align}
	[L_1,L_{-1}]\,=\,& \sum_{m>1,k>0}\bigg([b_{1-m}b_m, b_{-1-k}b_k] + [c_{1-m}c_m, c_{-1-k}c_k]    \bigg)-[b_{1\over 2} c_{1\over 2} , b_{-{1\over 2}} c_{-{1\over 2}}]\cr
	\,=\,&  2 b_{-{1\over 2}}b_{1\over 2} - 2 c_{-{1\over 2}}c_{1\over 2}   + 2 \sum_{m>1} (b_{-m}b_{m} - c_{-m} c_m)  - {1\over 4}\,=\, 2 L_0\ .
\end{align}
In addition, we also double-check the central charge by 
\begin{align}
	[L_2,L_{-2}]\,=\,& \sum_{m>2,k>0}\bigg([b_{2-m}b_m, b_{-2-k}b_k] + [c_{2-m}c_m, c_{-2-k}c_k]    \bigg)-[b_{1\over 2} c_{3\over 2} , b_{-{1\over 2}} c_{-{3\over 2}}]-[b_{3\over 2} c_{1\over 2} , b_{-{3\over 2}} c_{-{1\over 2}}]\ ,\cr
	\,=\,&  4 b_{-{1\over 2}}b_{1\over 2} - 4 c_{-{1\over 2}}c_{1\over 2} + 4 b_{-{3\over 2}}b_{3\over 2} - 4 c_{-{3\over 2}}c_{3\over 2}   + 4 \sum_{m>2} (b_{-m}b_{m} - c_{-m}c_m ) - {3\over 2}\,=\, 4 L_0 -1 \ .
\end{align}

\section{Appendix C: $\alpha$-Vacua}
\label{App: alpha vacua}

We begin with the $sl(2,\mathbb{R})$ generators defined by
\begin{align}
	J^0\,=\,&\sum_{n>0} {1\over 2|n|} (b_{-n} b_n + c_{-n}c_n)\ ,\\
	J^+\,=\,&\sum_{n>0} {1\over |n|} b_{-n} c_n \ ,\\
	J^-\,=\,&\sum_{n>0} {1\over |n|} c_{-n} b_n \ ,
\end{align}
where they satisfy the $sl(2,\mathbb{R})$ algebra
\begin{align}
	[J^0, J^\pm]\,=\,\pm J^\pm\;\;,\quad  [J^+,J^-]\,=\,-2 J^0\ .
\end{align}
Furthermore, they also commute with $L_0$
\begin{align}
	[L_0, J^0]\,=\,[L_0,J^\pm]\,=\,0 \ .
\end{align}
Note that the vacuum is invariant under the $SL(2,\mathbb{R})$ transformation, {\it i.e.}
\begin{align}
	J^0|0\rangle \,= \, J^\pm |0\rangle \,=\, 0\ .
\end{align}
Let us now consider the generator of Bogoliubov transformations:
\begin{align}
	\cG_\alpha\,=\, & i \sum_{n>0} {\alpha_n \over n} \big( b_{-n}c_{-n} + b_n c_n  \big)\ .
\end{align}
The Bogoliubov generator is Hermitian. However it is not $\cJ$-Hermitian; it flips the sign of all $\alpha$'s {\it i.e.}
\begin{align}
	\cG_\alpha^\dag \,=\, \cG_\alpha\;\;,\quad \cG_\alpha^{\dag_\alpha}\,=\, - \cG_{\alpha} \,=\, \cG_{-\alpha}\ .
\end{align}

The Bogoliubov transformation 
generated by $\cG_\alpha$ is not a unitary transformation but a similarity transformation that preserves canonical anti-commutation relations of the fermi oscillators
\begin{align}
	\tilde{b}_n^{(\alpha)}\,\equiv\, &e^{-i\cG_\alpha }b_n e^{i\cG_\alpha} \,=\,  \cosh \alpha_n b_n -\sinh \alpha_n c_{-n} \, \cr
	\tilde{c}_{-n}^{(\alpha)}\,\equiv\, &e^{-i\cG_\alpha }c_{-n} e^{i\cG_\alpha} \,=\,  -\sinh \alpha_n b_n +\cosh \alpha_n c_{-n}\ ,\cr
	\tilde{b}_{-n}^{(\alpha)}\,\equiv\, &e^{-i\cG_\alpha }b_{-n} e^{i\cG_\alpha} \,=\,  \cosh \alpha_n b_{-n} -\sinh \alpha_n c_{n}\ ,\cr
	\tilde{c}_n^{(\alpha)}\,\equiv\, &e^{-i\cG_\alpha }c_n e^{i\cG_\alpha} \,=\,  -\sinh \alpha_n b_{-n} + \cosh \alpha_n c_{n}\ . \label{eq: bogoliubov transf}
\end{align}
Note that the adjoint action of $\cJ$ on the $\tilde{b}_n^\alpha$ and $c_n^\alpha$ flips not only $n$ but also $\alpha$'s
\begin{align}
	\cJ b_n^{(\alpha)} \cJ \,=\, b_{-n}^{(-\alpha)}\;\;,\quad \cJ c_n^{(\alpha)} \cJ \,=\, c_{-n}^{(-\alpha)}\ .
\end{align}

We define the $\alpha$-vacuum which is annihilated by $\tilde{b}_n^{(\alpha)}$ and $c_n^{(\alpha)}$ ($n>0$)
\begin{align}
	|\alpha\rrangle \,=\,  {e^{-i \cG  }\over \sqrt{\cN} }|0\rangle\, =\, \prod_{n>0} \bigg( {\cosh \alpha_n  + \sinh\alpha_n {1\over n} b_{-n} c_{-n} \over \sqrt{\cosh 2\alpha_n}} \bigg) \; |0\rangle \ ,
\end{align}
where the normalization is chosen to be
\begin{align}
	\cN= \prod_{n>0} \cosh 2\alpha_n\ .
\end{align}
Here we consider only the holomorphic part for simplicity, 
and the calculations of the anti-holomorphic part are the same. 
The $\alpha$-vacuum bra state is given by
\begin{align}
	|\alpha \rrangle \,=\, {1\over \sqrt{\cN}} e^{-i \cG_\alpha} |0\rangle \quad\Longrightarrow \quad \llangle \alpha| \,=\,\langle 0| \big(e^{-i \cG_\alpha }\big)^{\dag_\cJ}{1\over \sqrt{\cN}}\,=\,\langle 0| e^{-i \cG_\alpha }{1\over \sqrt{\cN}}\ .
\end{align}
And one can confirm that the $\alpha$-vacuum is normalized
\begin{align}
	\llangle \alpha | \alpha \rrangle \,=\, {1\over \cN}\langle 0 | e^{-i \cG_\alpha} e^{-i\cG_\alpha} |0 \rangle \,=\,  {1\over \cN} \langle 0 | e^{-iG_{2\alpha}}  |0 \rangle\,=\,1\ .
\end{align}
Unlike the TFD state, $e^{-i\cG_\alpha}$ is not $\cJ$-unitary, which leads to non-trivial normalization constant $\cN$ for the $\alpha$-vacuum. The $\alpha$-vacuum is not invariant under the action of $\cJ$ but flips the sign of all $\alpha$'s, {\it i.e.}
\begin{align}
	\cJ\; |\alpha\rrangle \,=\,|-\alpha\rrangle\ .
\end{align}
Also note that the Bogoliubov generator commutes with $sl(2,\mathbb{R})$ generators
\begin{align}
	[\cG_\alpha,J^0]\,=\,0\;\;,\quad [\cG_\alpha,J^\pm]\,=\, 0\ .
\end{align}
Therefore the $\alpha$-vacuum is invariant under the $SL(2,\mathbb{R})$ transformation
\begin{align}
	J^0|\alpha\rrangle \,=\, J^\pm|\alpha\rrangle\,=\, 0 \ .
\end{align}

Now we evaluate the two point function function $G_\alpha(z,w)$ with respect to $\alpha$-vacuum
\begin{align}
	&G_\alpha(z,w)\,\equiv\,\llangle \partial \bar{\psi}(z)  \partial \psi (w)  \rrangle_{\alpha}  \,=\, -{1\over 4 \pi} \sum_{n,m>0} \langle \alpha|   \big(  b_{-n} z^{n-1}+ c_n z^{-n-1}  \big) \big( b_m w^{-m-1} +c_{-m} w^{m-1} \big) |\alpha \rangle\ ,\cr
	\,=\,& -{1\over 4\pi  }  \sum_{n>0} {1\over  \cosh 2\alpha_n} \langle 0| \big( \cosh \alpha_n + b_n c_n {1\over n} \sinh \alpha_n \big)  \big(  b_{-n} z^{n-1}+  c_n z^{-n-1} \big) \big( b_n w^{-n-1} +c_{-n} w^{n-1} \big) \big( \cosh \alpha_n  + \sinh\alpha_n {1\over n} b_{-n} c_{-n} \big) |0\rangle \ ,\cr
	\,=\,& {1\over 4\pi } {1\over zw} \sum_{n>0} n \bigg[ \bigg({w\over z}\bigg)^n {\cosh^2\alpha_n\over \cosh 2\alpha_n} - \bigg({z\over w}\bigg)^n {\sinh^2\alpha_n\over \cosh 2\alpha_n} \bigg]\ .
\end{align}
When we set all $\alpha_n$'s to be the same value $\alpha$, we find
\begin{align}
	G_\alpha(z,w)\,=\,\llangle \partial \bar{\psi}(z)  \partial \psi (w)  \rrangle_{\alpha_n=\alpha} \,=\,  {1\over 8\pi }  { \sqrt{z\over w}+ \sqrt{w\over z} \over (z-w)^2}\ ,
\end{align}
which is independent of $\alpha$, and it is identical to the two-point function with respect to the vacuum $|0\rangle$.

On the other hand, one may calculate the two-point function with naive norm. Note that the definition of $\alpha$-vacuum bra state is changed with naive norm, {\it i.e.} 
\begin{align}
	|\alpha \rangle \,=\, e^{-i \cG_\alpha} |0\rangle \quad\Longrightarrow \quad \langle \alpha| \,=\,\langle 0| \big(e^{-i \cG_\alpha }\big)^{\dag}\,=\,\langle 0| e^{i \cG_\alpha } \ ,
\end{align}
The two-point function with respect to $\alpha$-vacuum with naive norm is 
\begin{align}
	&\langle \partial \bar{\psi}(z)  \partial \psi (w)  \rangle_\alpha \,=\, -{1\over 4 \pi} \sum_{n,m>0} \langle \alpha|   \big( c_n z^{-n-1} + b_{-n} z^{n-1} \big) \big( b_m w^{-m-1} +c_{-m} w^{m-1} \big) |\alpha \rangle\ ,\cr
	\,=\,& -{1\over 4 \pi} \sum_{n,m>0} \langle 0|    e^{i\cG_\alpha } \big( c_n z^{-n-1} + b_{-n} z^{n-1} \big) \big( b_m w^{-m-1} +c_{-m} w^{m-1} \big)e^{-i\cG_\alpha } |0 \rangle\ ,\cr
	\,=\,& {1\over 4\pi z w}\sum_{n>0} n \bigg[ \bigg({w\over z}\bigg)^n\cosh^2\alpha_n+ \bigg({z\over w}\bigg)^n\sinh^2\alpha_n    + \big((zw)^n + (zw)^{-n} \big)\sinh \alpha_n \cosh\alpha_n   \bigg]\ .
\end{align}
Again, if one can take the same value of $\alpha_n$ for all $n$, then we have
\begin{align}
	&\langle \partial \psi^\dag(z)  \partial \psi (w)  \rangle_\alpha = {1\over 8\pi }  { \sqrt{z\over w}+ \sqrt{w\over z} \over (z-w)^2} \cosh(2\alpha) + {1\over 8\pi } {1\over w^2} { \sqrt{z w}+ \sqrt{1\over z w} \over \big(z-{1\over w}\big)^2} \sinh(2\alpha)\ ,\\
	\,=\,& G_0(z,w)\cosh^2\alpha + {1\over z^2 w^2} G_0(1/z,1/w)\sinh^2\alpha + {1\over 2}\sinh 2\alpha \bigg[{1\over z^2} G_0(1/z,w) + {1\over w^2} G_0(z,1/w)  \bigg]\ .
\end{align}
Note that the second term diverges when $zw=1$. This is analogous 
to the two point function $G^{dS}_{\alpha}(z,w)$ of the scalar field in the $\alpha$-vacua of the de Sitter space 
where the two-point function diverges for antipodal points, which is given by
\begin{align}
	G^{dS}_{\alpha}(z,w)\,=\, G^{dS}_E(z,w)\cosh^2\alpha + G^{dS}_E(z_A,w_A)\sinh^2\alpha + {1\over 2}\sinh 2\alpha \bigg[G^{dS}_0(z_A,w) + G^{dS}_0(z,w_A)  \bigg]\ ,
\end{align}
where $G^{dS}_E(z,w)$ is the two point function with respect to Bunch-Davis vacuum (i.e. $\alpha=0$) and $z_A$ is the antipodal point of $z$.

\section{Appendix D: Non-$\cJ$-Hermiticity and Real Eigenvalue}
\label{App: non hermiticity}

Instead of the mode expansion~\eqref{eq: mode expansion of psi} of $\psi$ and $\bar{\psi}$, one may expand $\psi$ and $\bar{\psi}$ in terms of $\tilde{b}^{(\alpha)}_n$ and $\tilde{c}^{(\alpha)}_n$ via Eqs.~\eqref{eq: bogoliubov transf}
\begin{align}
	\psi(z,\bar{z}) \,=\, & {i \over \sqrt{4\pi }} \sum_{n>0} {1\over n} \big(b_n z^{-n} - c_{-n} z^n + \bar{b}_n \bar{z}^{-n} - \bar{c}_{-n} \bar{z}^n\big)\ ,\cr
	\,=\,& {i \over \sqrt{4\pi }} \sum_{n>0} {1\over n} \bigg[ \big(\cosh \alpha_n \tilde{b}^{(\alpha)}_n + \sinh \alpha_n \tilde{c}^{(\alpha)}_{-n}\big) z^{-n} - \big( \sinh \alpha_n \tilde{b}^{(\alpha)}_n +\cosh \alpha_n \tilde{c}^{(\alpha)}_{-n}\big) z^n\cr
	& \hspace{20mm} + \big(\cosh \bar{\alpha}_n \tilde{\bar{b}}^{(\bar{\alpha})}_n +\sinh \bar{\alpha}_n \tilde{\bar{c}}^{(\bar{\alpha})}_{-n}\big)  \bar{z}^{-n} -  \big( \sinh \bar{\alpha}_n \tilde{\bar{b}}^{(\bar{\alpha})}_n +\cosh \bar{\alpha}_n \tilde{\bar{c}}^{(\bar{\alpha})}_{-n}\big)  \bar{z}^n\bigg)\ ,\label{eq: mode expansion alpha}
\end{align}
and similar for $\bar{\psi}$. Here $\bar{\alpha}_n$ ($n>0$) is the Bogoliubov transformation parameter for the anti-holomorphic part. Recall that the operator $\cJ$ flips the sign of $\alpha_n$ and $\bar{\alpha}_n$ in the oscillators. Therefore it is convenient to define $\tilde{\cJ}$ operator which is different from the $\cJ$ operator
\begin{align}
	\tilde{\cJ}\,=\, \exp\bigg[ \pi i \sum_{n>0} {1\over n} \big(\tilde{c}_{-n}^{(\alpha)}\tilde{c}_n^{(\alpha)} + \tilde{\bar{c}}_{-n}^{(\alpha)}\tilde{\bar{c}}_n^{(\alpha)} \big)\bigg]\,\ne \, \cJ\ .
\end{align}
One can repeat the same procedure with the operator $\tilde{\cJ}$ instead of $\cJ$ ({\it e.g.} $\tilde{\cJ}$-norm {\it etc.}). Then one may drop the tilde and $\alpha$ superscript in the oscillators and the operator $\tilde{J}$. Or from the beginning, one may use the non-trivial mode expansion \eqref{eq: mode expansion alpha} to quantize the oscillators.

With the mode expansion~\eqref{eq: mode expansion alpha} (with tilde and $\alpha$ dropped), $L_0$ is given by
\begin{align}
	L_0 = \sum_{n>0} \big[ &\cosh 2\alpha_n (b_{-n} b_n - c_{-n}c_n) + \sinh 2\alpha_n (b_{-n}c_{-n} + c_n b_n) -2n \sinh^2\alpha_n \big]\ .
\end{align}
up to the vacuum energy density $-{1\over 8}$. Note that $L_0$ is not $\cJ$-Hermitian. 
Therefore it is not guaranteed that 
$L_0$ has real eigenvalues. To diagonalize $L_0$, 
it is sufficient to consider a subspace that consists of
\begin{align}
	|0\rangle\;\;,\quad |1\rangle \,\equiv\, {1\over \sqrt{n}}\, b_{-n}|0\rangle\;\;,\quad |2\rangle \,\equiv\, {1\over \sqrt{n}} \, c_{-n}|0\rangle\;\;,\quad |3\rangle \,\equiv\, {1\over n} \,b_{-n} c_{-n} |0\rangle \ .
\end{align}
The matrix element $M_{jk}$ of the operator $L_0$ with double-bracket is given by
\begin{align}
	M_{ji}\,\equiv\, \llangle j |\;  L_0 \; | k \rrangle \hspace{10mm} (j,k=0,1,2,3)\ .
\end{align}
and we have
\begin{align}
	M\,=\,\begin{pmatrix}
	-2n \sinh^2 \alpha_n & 0 & 0& -n\sinh 2\alpha_n \\
	0 & n & 0 & 0 \\
	0 & 0 & n & 0 \\
	n\sinh 2\alpha_n & 0 & 0 & 2n\cosh^2\alpha_n  \ \\
	\end{pmatrix}\ .
\end{align}
The eigenvalues of the matrix $M$ are $\{0,n,n,2n\}$ which is identical to the eigenvalues of $L_0$ with the original mode expansion in~\eqref{eq: mode expansion of psi}. Although the operator is non-$\cJ$-Hermitian, it has real eigenvalues. This is because there exists a Bogoliubov transformation which makes the operator $L_0$ $\cJ$-Hermitian with new definition of $\cJ$ operator.

\end{widetext}

\bibliography{symplectic_arxiv}

\end{document}